\documentclass[sigplan,screen,10pt]{acmart}

\startPage{1}

\AtBeginDocument{%
  \providecommand\BibTeX{{%
    \normalfont B\kern-0.5em{\scshape i\kern-0.25em b}\kern-0.8em\TeX}}}

\setcopyright{acmcopyright}
\acmPrice{15.00}
\acmDOI{10.1145/3520312.3534868}
\acmYear{2022}
\copyrightyear{2022}
\acmSubmissionID{pldiws22mapsmain-p11-p}
\acmISBN{978-1-4503-9273-0/22/06}
\acmConference[MAPS '22]{Proceedings of the 6th ACM SIGPLAN International Symposium on Machine Programming}{June 13, 2022}{San Diego, CA, USA}
\acmBooktitle{Proceedings of the 6th ACM SIGPLAN International Symposium on Machine Programming (MAPS '22), June 13, 2022, San Diego, CA, USA}

\usepackage{alltt}
\usepackage{enumitem}
\usepackage{listings}
\usepackage{xspace}
\usepackage{multirow}
\usepackage{wrapfig}

\usepackage{booktabs}   
\usepackage{subcaption} 

\usepackage{etoolbox}

\newtoggle{extendedversion}

\definecolor{keyword}{HTML}{37AC4A}
\definecolor{operator}{HTML}{A51DFF}
\definecolor{string}{HTML}{C03333}
\definecolor{background}{HTML}{F7F7F7}
\definecolor{diffminus}{HTML}{FFEEF0}
\definecolor{diffplus}{HTML}{E6FFED}

\lstdefinelanguage{python}{
  columns=flexible,
  morestring=[b]',
  morestring=[b]",
  morestring=[b]""",
  stringstyle=\textcolor{string},
  showstringspaces=false,
  morecomment=[l]\#,
  commentstyle=\color{gray}\it\ttfamily,
  morekeywords={and,as,assert,break,class,continue,def,del,elif,else,except,False,finally,for,from,global,if,import,in,is,lambda,None,nonlocal,not,or,pass,raise,return,True,try,while,with,yield},
  keywordstyle=\color{keyword}\bf\ttfamily,
  otherkeywords={|,>>,\&,=,**},
  morekeywords=[2]{|,>>,\&,=,**},
  keywordstyle=[2]\color{operator}\bf\ttfamily,
}

\newcommand*{\python}[1]{\lstinline[language=python]{#1}}
\newcommand*{\pyplain}[1]{{\fontsize{7.2}{8.5}\texttt{#1}}}

\newcommand{\tool}{Maro\xspace}

\toggletrue{extendedversion}

\begin{document}

\title[Automatically Debugging AutoML Pipelines using Maro (Extended Version)]{Automatically Debugging AutoML Pipelines using\\ Maro: ML Automated Remediation Oracle (Extended Version)}

\author{Julian Dolby}
\affiliation{
  \institution{IBM Research}            
  \country{USA}                    
}
\email{dolby@us.ibm.com}         

\author{Jason Tsay}
\affiliation{
  \institution{IBM Research}            
  \country{USA}                    
}
\email{jason.tsay@ibm.com}         

\author{Martin Hirzel}
\affiliation{
  \institution{IBM Research}            
  \country{USA}                    
}
\email{hirzel@us.ibm.com}         

\begin{abstract}
  Machine learning in practice often involves complex pipelines for data
cleansing, feature engineering, preprocessing, and prediction.
These pipelines are composed of operators, which have to be correctly
connected and whose hyperparameters must be correctly configured.
Unfortunately, it is quite common for certain combinations of
datasets, operators, or hyperparameters to cause failures.
Diagnosing and fixing those failures is tedious and error-prone and
can seriously derail a data scientist's workflow.
This paper describes an approach for automatically debugging an ML
pipeline, explaining the failures, and producing a remediation.
We implemented our approach, which builds on a combination of AutoML
and SMT, in a tool called \tool.
\tool works seamlessly with the familiar data science ecosystem including
Python, Jupyter notebooks, scikit-learn, and AutoML tools such as Hyperopt.
We empirically evaluate our tool and find that for most cases, a single remediation automatically fixes errors, produces no additional faults, and does not significantly impact optimal accuracy nor time to convergence.

\end{abstract}




\maketitle

\section{Introduction}

Artificial Intelligence (AI) is an exciting rising paradigm of
software development that however also comes with many new challenges
for developers.
Challenges range from systemic issues such as a lack of education and
training~\cite{amershi2019} and difficulty in
reproducibility~\cite{gundersen2017} to hidden technical
debt~\cite{sculley2015} to a need for fairness and controlling for
bias~\cite{bellamy2019}.
Individual AI developers developing software that trains machine
learning (ML) models face tasks covering a wide range from data
collection and cleaning to feature selection to training and
evaluating models.
These tasks are often highly entangled, where errors in earlier tasks
often have serious or insidious cross-cutting
consequences~\cite{hill2016}.
Consequences of errors span a wide range depending on the components
that they affect, from hard faults to data corruption to incorrect or
unintended functionality in the AI system~\cite{islam2019}.
Similarly, the potential causes of errors are numerous, from the
dataset used, derived features, hyperparameters, operators, etc.
This complexity in reasoning and tracking errors in AI systems makes
them difficult for AI developers to debug.

This paper focuses on the task of debugging a set of possible ML
pipelines for a given dataset.
Following the terminology of scikit-learn~\cite{scikit-learn}, a
popular ML framework, we define an ML \emph{pipeline} as a graph of
operators and their hyperparameters.
Once trained, an ML pipeline becomes an ML model that supports
evaluation using metrics and predictions on new unseen data.
For this work, we consider \emph{planned pipelines}, which specify a
graph of ML operators and schemas for hyperparameters, but leave some
choices open, such as concrete hyperparameter settings, or picking one
of a choice of multiple operators at a given pipeline step.
Given a planned pipeline, a \emph{pipeline instance} fills in all the
choices, by picking operators from the set of available options and
hyperparameter values from the domain of the corresponding schema.
It is common practice to use an automated machine learning
(\emph{AutoML}) tool to explore the search space of choices in a
planned pipeline to find the best pipeline instance for a given dataset.
A pipeline instance is trainable, and can thus be turned into a model
and evaluated against metrics for a given dataset.
An AutoML search generates and evaluates multiple pipeline instances.

We focus on debugging these planned pipelines because errors in them often propagate to the derived models. Additionally, the automated search often tries erroneous combinations of operators and hyperparameters, which is wasteful. Debugging the failures of a particular ML pipeline is difficult and time-consuming due to the experimental nature of AI development along with the multitude of possible failure causes~\cite{arpteg2018}. Often, the lack of transparency and explainability in AI development results in developers treating pipelines as ``black boxes,'' forcing a trial-and-error approach of testing by running models repeatedly~\cite{hill2016}. This is combined with a difficulty of localizing the error due to entanglement or hidden feedback loops~\cite{sculley2015}. Rather than reason about the development process as a whole with all of its complexities when debugging, our tool embraces the iterative nature of AI development to more efficiently find and remediate bugs.

Our approach
combines automated machine learning (AutoML) with a satisfiability
modulo theories (SMT) solver to generate, analyze, and remediate
instances of a planned ML pipeline for a given task.
The complexity and sheer amount of possible causes of failure makes
manual debugging difficult~\cite{nushi2017}.
With AutoML, the amount of experiments to reason across when debugging
only increases.
Our system eases this burden on the AI developer by viewing debugging
as a search for constraints over a given space of operators
and their hyperparameters, which is a natural fit for an SMT solver.
Thus, our system attempts to automatically determine which
constraints of operators or hyperparameters prevent certain failures.
By using these constraints and the original planned ML pipeline, we
generate a remediated planned pipeline that avoids (a generalization
of) these failures.

This paper presents a tool named \tool (ML Automated Remediation
Oracle) that automatically debugs ML pipelines and generates
remediated pipelines based on AutoML experiment results. We build on
top of a Python-based open source AutoML interface named
Lale~\cite{baudart_et_al_2020,baudart_et_al_2021} that supports composing operators from
popular ML libraries such as scikit-learn~\cite{scikit-learn} into
pipelines and then running AutoML optimizers such as
Hyperopt~\cite{hyperopt} across these pipelines. Given a user's ML
pipeline and their initial AutoML-generated experiments, if some
of the experiments have failed, then \tool automatically returns a
remediated pipeline. Our tool also provides
explanations for the automated remediations for the given ML pipeline
through rendering the constraints found by the solver in natural language,
as well as displaying the differences between the original and remediated
pipelines. We evaluate \tool on 20 planned pipelines that cover a diverse
set of ML operators, failure causes, and remediation requirements.
We compare \tool against approaches from the BugDoc pipeline
debugger~\cite{lourenco2020}.
Since BugDoc does not provide remediation, we extend it with this
feature to enable better experimental comparison.
To the best of our knowledge, our tool is the first to provide a full
debugging and remediation round-trip.

The contributions of \tool are as follows:
\begin{enumerate}[leftmargin=\parindent]
  \item{An approach for automated fault localization in ML pipe\-lines based on AutoML and SMT solvers.}
  \item{Automated remediation for ML pipelines, by applying constraints found by the localizer to the original pipeline.}
  \item{Explanation of remediations via natural language as well as via differencing the original and remediated pipeline.}
\end{enumerate}

\section{Overview and Examples}\label{sec:tooldescription}

This section uses examples to give a high-level description of our
tool. The target persona is Dante, a data scientist. Dante uses
popular Python machine-learning libraries from a Jupyter notebook to
build predictive models.

\subsection{Detailed Example}\label{sec:detailed_example}

Our first example starts when Dante has already inspected the data and
found that
it has some missing values, categorical features, and discrete target
labels.  So he assembles a \python{planned} pipeline with
three steps: a \python{SimpleImputer} for filling in missing values, a
\python{OneHotEncoder} for transforming categoricals into numbers, and a
\python{LogisticRegression} classifier for predicting target labels.
The pipe combinator (\python{>>}) connects operators with
dataflow edges, creating a pipeline.

\begin{lstlisting}[language=python,frame=single,framerule=0.5pt,backgroundcolor=\color{background},belowskip=3mm, aboveskip=3mm,xleftmargin=2pt, xrightmargin=2pt]
one_hot_encoder = OneHotEncoder(handle_unknown="ignore")
planned = SimpleImputer >> one_hot_encoder >> LogisticRegression
\end{lstlisting}

Dante's day-to-day workflow involves
trial-and-error with different pipelines to find
the best-performing one. Rather than doing all experiments by hand,
Dante uses AutoML tools to automate some of that search. In the
example, both \python{SimpleImputer} and \python{LogisticRegression}
have hyperparameters that Dante deliberately left unspecified.
Instead, he uses \python{Hyperopt}~\cite{hyperopt} to search possible
configurations for them, based on hyperparameter schemas specified in
the library.
Each evaluation picks a pipeline instance
(a pipeline where all hyperparameters are bound to values drawn from
their schema) and evaluates it using cross-validation.

\begin{lstlisting}[language=python,frame=single,framerule=0.5pt,backgroundcolor=\color{background},belowskip=2mm, aboveskip=3mm,xleftmargin=2pt, xrightmargin=2pt]
hyperopt_trainable = Hyperopt(estimator=planned, max_evals=20)
hyperopt_trained = hyperopt_trainable.fit(train_X, train_y)
\end{lstlisting}
\begin{lstlisting}[frame=single,framerule=0pt,aboveskip=0mm, belowskip=1mm]
Please wait ...
Done, 15 out of 20 evaluations failed, call summary() for details.
\end{lstlisting}

Unfortunately, most evaluations failed, i.e., the corresponding
pipeline instance raised an exception. Dante wonders what he should
do now. He is tempted to just ignore the failures and move on, but
what if there are root causes that he should
understand to build a better pipeline? Given how many evaluations
failed, the search may be less effective, as it covered less ground.
Moreover, the failures do not come for free: they may have wasted
computational resources before raising their exceptions. So rather
than give in to the temptation, he decides to poke around a bit. But
that prospect fills him with dread: it can become a time drain,
since comparing even a moderate number of pipeline instances (like 20 in this
example) is tedious.  For now, Dante decides to at least call the
\python{summary()} method as suggested by the error message.

\begin{lstlisting}[language=python,frame=single,framerule=0.5pt,backgroundcolor=\color{background}, aboveskip=3mm, belowskip=2mm,xleftmargin=2pt, xrightmargin=2pt]
hyperopt_trained.summary()
\end{lstlisting}
\includegraphics[width=0.55\columnwidth]{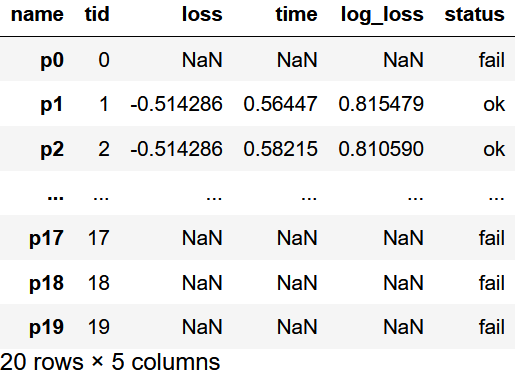}

Each evaluation in the summary has a name, ID, loss (in this case
accuracy, negated to make it a minimization problem), log-loss, and
status. Dante decides to retrieve
one of the failing instances and pretty-print it as Python code.

\begin{lstlisting}[language=python,frame=single,framerule=0.5pt,backgroundcolor=\color{background},xleftmargin=2pt, xrightmargin=2pt]
hyperopt_trained.get_pipeline("p0").pretty_print()
\end{lstlisting}
\begin{lstlisting}[language=python, aboveskip=0mm, belowskip=2mm,xleftmargin=2pt, xrightmargin=2pt]
simple_imputer = SimpleImputer(strategy="median")
one_hot_encoder = OneHotEncoder(handle_unknown="ignore")
logistic_regression = LogisticRegression(
    dual=True,
    fit_intercept=False,
    intercept_scaling=0.48518719297596336,
    max_iter=326,
    solver="liblinear",
    tol=0.006373368408152854,
)
pipeline = simple_imputer >> one_hot_encoder >> logistic_regression
\end{lstlisting}

As expected, \python{Hyperopt} chose concrete hyperparameters.  But
what went wrong? Dante could now look at all the other pipeline instances to
find out which choices cause failures. Or he could try to train them
and wade through their exception back-traces. Instead, Dante asks
\tool, the tool introduced by this paper, for guidance. \tool has three
parts: a fault localizer, a remediator, and an explainer.  The
\python{auto_remediate()} function first calls the fault localizer and the
remediator, taking the original planned pipeline and the
evaluations from the Hyperopt run (pipeline instances and their
status) and returning a new \python{remediated} pipeline. The
\python{remediated} pipeline is as similar as possible to the original
\python{planned} pipeline while ruling out all failures observed in
earlier evaluations. Lastly, the \emph{explainer} returns a natural language
explanation of the suggested remediation.

\begin{lstlisting}[language=python,frame=single,framerule=0.5pt,backgroundcolor=\color{background},xleftmargin=0.5pt, xrightmargin=0.5pt]
remediated = auto_remediate(planned, hyperopt_trained,
  explanation=True)
\end{lstlisting}

\begin{lstlisting}[frame=single,framerule=0pt,aboveskip=1mm,belowskip=2mm]
Try setting argument 'strategy' in operator SimpleImputer to 'most_frequent'
\end{lstlisting}
%

The explanation pinpoints the cause of failure:
\python{SimpleImputer} should use the \python{"most_frequent"}
strategy. This makes sense, since the dataset is categorical, and
other imputation strategies (such as \python{"median"}) require
numeric data. Dante is relieved that \tool guided him to a solution,
and decides to try out the remediated pipeline.
The remediated pipeline is again a planned pipeline for which
Hyperopt tries pipeline instances by searching the remaining
hyperparameters.

\begin{lstlisting}[language=python,frame=single,framerule=0.5pt,backgroundcolor=\color{background},xleftmargin=2pt, xrightmargin=2pt]
hyperopt_trainable = Hyperopt(estimator=remediated, max_evals=20)
hyperopt_trained = hyperopt_trainable.fit(train_X, train_y)
\end{lstlisting}
\begin{lstlisting}[frame=single,framerule=0pt,aboveskip=0mm,belowskip=2mm]
Please wait ...
Done, all evaluations succeeded.
\end{lstlisting}

This time, all 20 out of 20 evaluations succeeded. So Dante can get
back to his work of finding the best pipeline for the dataset. He can
evaluate the pipeline on test data, or perhaps use AutoML to search
different classifier choices.

\subsection{Tool Overview}\label{sec:overview}

\begin{figure}
\centerline{\includegraphics[width=\columnwidth]{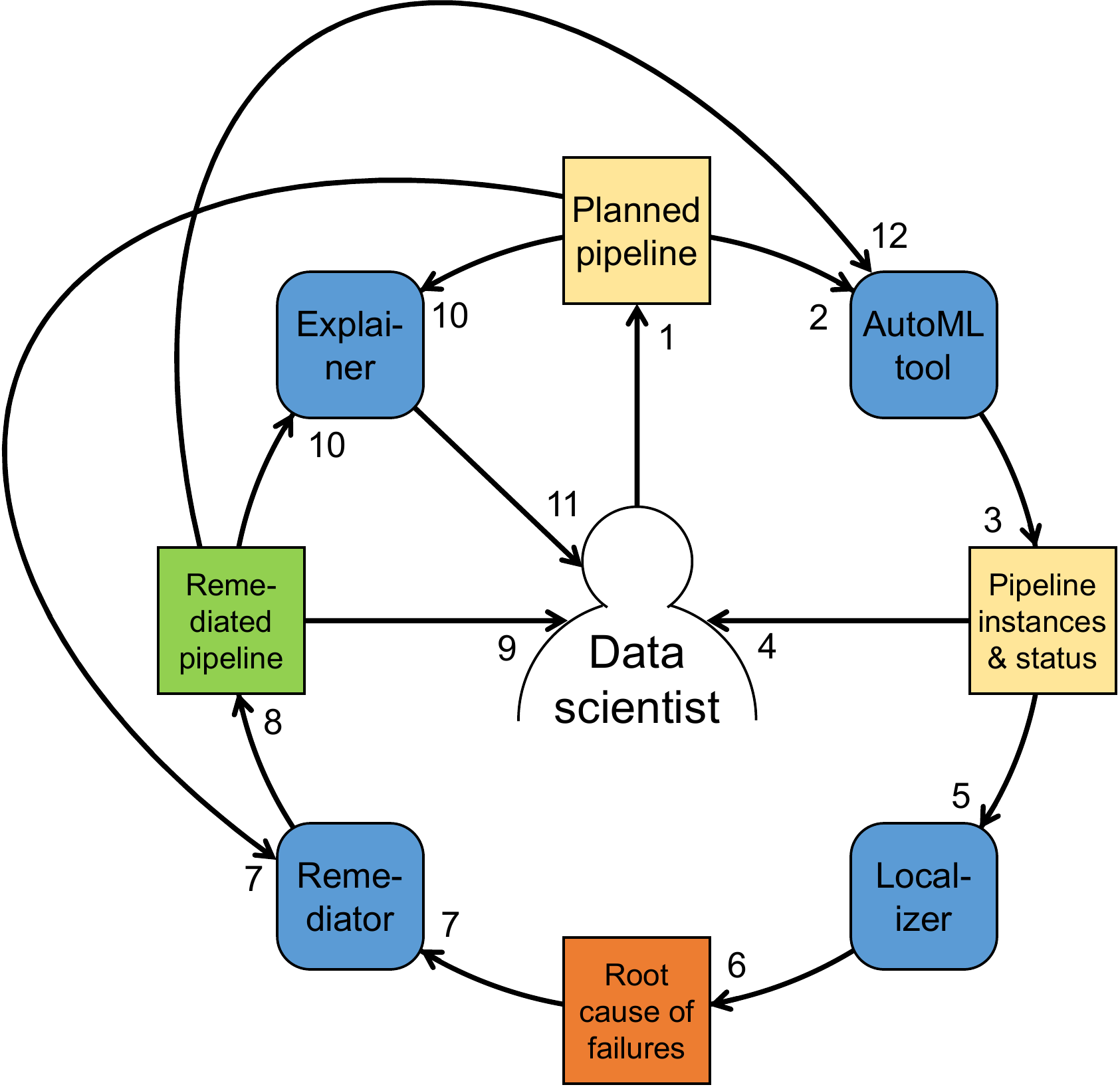}}
\caption{\label{fig:overview}Iterative ML development with \tool.}
\vspace{-5mm}
\end{figure}

Figure~\ref{fig:overview} gives an overview of how a data scientist
such as Dante can use our tool \tool. The workflow starts with the
data scientist, shown in the center, creating a planned pipeline~(1).
They can then feed this pipeline to an AutoML tool, such as grid-search, Hyperopt, or any other backends that Lale supports~(2). The
automated search yields a set of pipeline instances along with their
status, which can be ``ok'' or ``fail''~(3). Without \tool, the data
scientist would have little choice but to manually inspect these
results~(4). But a better option is to send the results on to \tool's
fault localizer component~(5). The localizer uses an SMT solver to find a root
cause of the failures~(6).  This root cause, along with the original
planned pipeline, forms the input to \tool's remediator
component~(7). The result is a remediated pipeline~(8), which the data
scientist can inspect directly if they so wish~(9). Alternatively, to
make the fix easier to understand, the data scientist can send the
remediated pipeline and the original pipeline to \tool's explainer
component~(10). This explains the remediation to the data scientist by
rendering it in natural language~(11).
And finally, as the remediated pipeline is itself a planned pipeline,
the data scientist can use it as input to the AutoML tool~(12),
thus completing the circle.

\subsection{Additional Use Cases}\label{sec:examples}

\tool can handle a diverse set of ML pipelines and associated failures.
This paper experiments with a set of 20 planned pipelines.
We initially chose a set of pipelines based on interviewing ML
practitioners and analyzing publicly-available pipelines.
Then, we grew that set as we implemented and tested \tool to exercise
challenging corner cases.
All planned pipelines use common ML operators, mostly from
scikit-learn~\cite{scikit-learn}, such as \python{LogisticRegression},
or \python{OneHotEncoder}, but also operators
from other scikit-learn compatible libraries, such as a bias
mitigator from AIF360~\cite{bellamy2019} and gradient-boosted trees
from LightGBM~\cite{ke_et_al_2017}.
\iftoggle{extendedversion}{
  The full list of pipelines is available in Table~\ref{fig:examples} in the appendix.
}{
  The full list of pipelines is available in the extended version of this paper~\cite{dolby2022automatically}.
}

The pipelines failed for a variety of reasons, including
characteristics of the input data;
incompatible operators;
incompatible hyperparameters;
or some combination of the above.
Some pipelines failed fast, others only after expensive training of a prefix.
Sometimes, even hyperparameters within a single operator can be
incompatible with each other.  This is known as a conditional
hyperparameter constraint, and some AutoML tools prune invalid
combinations from the search space based on manual specification,
e.g., auto-sklearn~\cite{feurer2015} or Lale~\cite{baudart_et_al_2021}.
However, other AutoML tools do not come with comprehensive conditional
hyperparameter constraint specifications, e.g., scikit-learn's
GridSearchCV.

\tool repairs each planned
pipeline in our set to prune failing instances from the search space.
Remediations may involve
removing operators from choice for AutoML algorithm selection;
limiting categorical hyperparameters to a set of values (such as the
complement of removing a value from an enum);
placing upper or lower bounds on continuous hyperparameters;
or some combination of the above.
While data is sometimes but not always part of the problem,
remediation is always in the pipeline, not in the data.
This is because in practice, data scientists must work with the data
at hand.
Thankfully, often, the purpose of an operator is to transform the data
you have into the data you need, so picking and configuring operators
in a pipeline can also fix data problems.

\section{Algorithms and Tool Design}\label{sec:system}

As shown in Figure~\ref{fig:overview}, \tool has three main components:

\begin{enumerate}[leftmargin=\parindent]
\item A \emph{localizer} that, given a set of evaluations, computes a
  root cause of failures, i.e., operator choices and hyperparameter
  settings that correlate with pipeline instances that failed.
\item A \emph{remediator} that, given the original planned pipeline
  and the root cause of failures, constructs a new
  pipeline that excludes known failures while allowing other settings.
\item An \emph{explainer} that, given the original planned pipeline
  and the root cause of failures, computes an explanation that makes the
  remediation easier to understand.
\end{enumerate}
We start with some preliminaries and defining \tool's interfaces,
then present how the three main components work.

\subsection{Preliminaries}\label{sec:preliminaries}

The input to \tool consists of a set of evaluations, which are
pipeline instances along with their status and the
original planned pipeline.

\begin{definition}[Pipeline]
  A planned \emph{pipeline} $P$ is a set of steps $S_0, \dots, S_n$,
  which are operators or operator choices.
\end{definition}

\begin{definition}[Pipeline instance]
  A \emph{pipeline instance} $p$ is a pipeline along with
  a Boolean result $r_p$ denoting success or failure and
  a mapping $H_p$ from hyperparameters to values.
\end{definition}

To simplify the discussion, we model operator choice for
algorithm selection by the presence of a hyperparameter
that identifies the chosen operator.

\paragraph{Constraints}

\tool uses an interface of {\em constraints} to communicate between
the fault localizer and the remediator: the localizer computes
con\-straints that capture successful runs, and the remediator alters
the initial planned pipeline to rule out pipeline instances that
violate those constraints.

\begin{figure}
\begin{lstlisting}[language=python,frame=single,framerule=0.5pt,backgroundcolor=\color{background},xleftmargin=2pt, xrightmargin=2pt]
one_hot_encoder = OneHotEncoder(handle_unknown='ignore')
ordinal_encoder = OrdinalEncoder(handle_unknown='ignore')
encoder_choice = one_hot_encoder | ordinal_encoder
planned = (project_categoricals >> encoder_choice
           >> StandardScaler >> LogisticRegression)
\end{lstlisting}
\vspace{-5mm}
\caption{Example pipeline (k).}
\label{fig:expl-k}
\vspace{-2mm}
\end{figure}

\begin{figure}
\begin{centering}
\[\begin{array}{l}
  \textrm{if }H_p(\pyplain{StandardScaler.with\_mean})=\pyplain{False}\\
  \qquad\textrm{then }\pyplain{True}\\
  \qquad\textrm{else }H_p(\pyplain{OrdinalEncoder.handle\_unknown})=\pyplain{"ignore"}
\end{array}\]
\end{centering}
  \vspace*{-4mm}
  \caption{Localizer-generated constraint for pipeline (k).}
  \label{fig:ex-fig}
  \vspace{-2mm}
\end{figure}

There are two kinds of constraints, atomic and multiple.
An atomic constraint compares a hyperparameter against a constant
(e.g., \mbox{$H_p(\pyplain{SimpleImputer.strategy})\neq\pyplain{"median"}$})
or against another hyperparameter or checks if a hyperparameter is present.
A multiple constraint arranges other constraints in an if-then-else tree.
To make this concrete, consider Figure~\ref{fig:ex-fig}, which shows
the constraints our solver found for example pipeline~(k), shown in
Figure~\ref{fig:expl-k}.
The if-part represents the top of the tree, checking whether
\pyplain{StandardScaler.with\_mean} is \pyplain{False}.
The then-clause is simply \pyplain{True}, indicating that the pipeline
is valid.
The else-clause says that otherwise, the pipeline is valid if
\pyplain{OrdinalEncoder.ignore\_unknown} is present and set to \pyplain{"ignore"},
implying that the operator choice picked \pyplain{OrdinalEncoder}.

\subsection{Fault Localization}\label{sec:localization}

\tool receives a set of pipeline instances $\mathfrak{P}$,
$p_1,\dots,p_n$, and computes hyperparameter constraints $C$ in the
format of Section~\ref{sec:preliminaries} that determine if a pipeline
instance fails. To allow the most flexibility
in determining the constraints while also ensuring that
remediation is feasible, our approach uses templates of constraints we
can handle but these templates are made flexible with symbolic
variables that control the specific constraints.

To do this, \tool uses the solver-aided language
Rosette~\cite{torlak_bodik_2013}.  Solver-aided languages allow programming with symbolic values.
 Intuitively, symbolic values can be used for any program value (of a
 supported type),
 and the result of running such a program is a logical formula that,
 when solved, yields concrete values for the given symbolic ones such
 that the program succeeds.
 This allows us to write logic that checks whether a given constraint explains
 all failures, leaving the actual constraint symbolic so that the solver
 fills it in.

 \paragraph{Atomic Constraint.}   To see how this works, consider the example from
 Section~\ref{sec:detailed_example},  where \pyplain{SimpleImputer} with
 hyperparameter \pyplain{strategy} set to \pyplain{"median"} breaks on non-numeric
 data.  This is an atomic constraint that invalidates pipelines, which
 is the simplest case.  If we somehow knew the constraint
 to use, we could write the following:

$E(\mathfrak{P}) \equiv
  \forall_{p \in \mathfrak{P}}
  \left(\begin{array}{l}
  r_p \iff\\
  H_p(\pyplain{SimpleImputer.strategy})\neq\pyplain{"median"}
  \end{array}\right)$

This formula states that a pipeline instance from
$\mathfrak{P}$ succeeds if and only if it does not bind
\pyplain{SimpleImputer.strategy} to \pyplain{"median"}.  If we think of
$E$ as instrumenting execution of AutoML, so it sees all attempted
pipelines and their outcomes, it will be true for the
example from Section~\ref{sec:detailed_example},  since those
pipelines indeed fail in precisely that case.
This would be simple to do; however, we do not, in general, know in
advance what hyperparameter to check.  But symbolic
variables---denoted by @ in Rosette---allow us to leave the actual
constraint unspecified and have the solver fill it in.  We can write
that as follows:$$S_1\left(\mathfrak{P}\right) \equiv \forall_{p \in \mathfrak{P}} \left( r_p
    \iff H_p(@\mathit{hparam}) = @\mathit{value}\right)$$

The process of abstractly executing the symbolic program plays the
role of the instrumentation mentioned above: the solver at the end
finds a binding of the symbolic variables that make execution valid,
if such there be.  Thus it binds the symbolic variables
$@\mathit{hparam}$ and $@\mathit{value}$ to concrete values
that make the assertion true.  This will find any hyperparameter
setting that correlates exactly with pipelines that fail.  In fact,
this simple logic suffices for any failure caused by a single value of
a single hyperparameter.  The symbolic variables can be read directly
from the solution to generate atomic constraints as described in
Section~\ref{sec:preliminaries}.

\iftoggle{extendedversion}{
There are several categories of error that similar constraints can
capture (discussed in more detail in the appendix).
}{
There are several categories of error that similar constraints can
capture (discussed in more detail in~\cite{dolby2022automatically}).
}
 They are the following ($S_2$, $S_3$, and $S_4$):
 \begin{itemize}[leftmargin=\parindent]
 \item presence of a hyperparameter, regardless of value
 \item numerical restriction to be more or less than a given value
 \item numerical constraints between hyperparameters
 \end{itemize}

\paragraph{Multiple Constraints.} While in some cases a single atom\-ic
 constraint suffices, that is not always the case.
Consider the example pipeline~(k), in which the combination of
\python{with_mean} for \python{StandardScaler} and
\python{handle_unknown} for \python{OneHotEncoder} breaks for this
dataset.
Either is allowed, but they cannot be used together.  To handle this, we stack
these constraints such that one constraint controls which other constraint
applies; the superscripts on $S$ indicate that the three uses of $S$
generate distinct symbolic variables, so there are three independent
constraints:$$
  S_5 \equiv \forall_{p \in \mathfrak{P}} \left(
  \begin{array}{l}
  r_p \iff\\
  \textrm{if }    S_\mathit{any}^1(\left\{p\right\})
  \textrm{ then } S_\mathit{any}^2(\left\{p\right\})
  \textrm{ else } S_\mathit{any}^3(\left\{p\right\})
  \end{array}\right)
$$
This is a tree structure of the constraints,
and the values for all the constraints can be read directly from the
variables produced by the solver.
This only illustrates two levels, but clearly they can be
stacked as deeply as needed.
The localizer communicates its results to the remediator by providing the
symbolic constraint $C_i$ of each $S^i$ in the format described in
Section~\ref{sec:preliminaries}, as exemplified in Figure~\ref{fig:ex-fig}.


\subsection{Remediation}\label{sec:remediation}

The remediator computes a remediated planned pipeline corresponding to
the formula $\textit{origPipe}\wedge C$, describing
a set of possible pipeline instances for AutoML to sample from.
Here, \textit{origPipe} is a formula that describes the original
planned pipeline.
It characterizes a (usually unbounded) set of possible pipeline instances
from which the initial AutoML run sampled a finite set of instances.
And $C$ is a formula returned by the localizer
that rules out a generalization of the concrete failed instances,
abstracted to be brief and broadly applicable.

As discussed in Section~\ref{sec:preliminaries}, the $C$ formula can involve
if-then-else, expressible via negation, conjunction, and disjunction.
Hence, one approach would be to perform remediation in a purely
logical representation and then, only at the end, convert back to a
pipeline representation suitable for AutoML tools.
Unfortunately, this would make the result of remediation inscrutable
for data scientists, since it may look nothing like \textit{origPipe}.
Therefore, for the sake of better explainability, \tool's remediation
algorithm takes a bottom-up approach of directly constructing a
remediated pipeline that resembles \textit{origPipe}.

\begin{figure}
\begin{lstlisting}[language=python,mathescape=true,morekeywords={algorithm,case},numbers=left,xleftmargin=5mm]
algorithm process(origPipe, $C$):
    case $C\;\equiv\;(\textrm{if }C_1\textrm{ then }C_2\textrm{ else }C_3)$:
        thenPipe = process(origPipe, $C_1 \wedge C_2$)
        elsePipe = process(origPipe, $\neg C_1 \wedge C_3$)
        return makeChoice(thenPipe, elsePipe)
    case $C\;\equiv\; C_1 \wedge C_2$:
        leftPipe = process(origPipe, $C_1$)
        return process(leftPipe, $C_2$)
    case isAtomicConstraint($C$):
        if affectsPresenceOfOperators($C$):
            tmpPipe = restrictChoice(origPipe, $C$)
        else:
            tmpPipe = origPipe
        if comparesMultipleHyperparameters($C$):
            return makeComparison(tmpPipe, $C$)
        else:
            return customizeSchemas(tmpPipe, $C$)
\end{lstlisting}
\vspace{-2mm}
\caption{\label{fig:rem_exp_algo}Pseudo-code for \tool's remediation algorithm.}
\vspace{-6mm}
\end{figure}

Figure~\ref{fig:rem_exp_algo} shows \tool's remediation algorithm.
As described in Section~\ref{sec:preliminaries}, the solver returns
constraints arranged as a tree, encoding conditionals where the parent
is an if-clause and subtrees represent then and else clauses.
\mbox{Lines 2--5} handle this case by recursive remediation calls for
the left and right subtree.
The \python{makeChoice} function combines the results via Lale's
choice combinator (\python{|}).
When the algorithm reaches a leaf, it faces a conjunction constraint,
handled by \mbox{Lines 6--8} via recursive calls to remediate conjuncts
one by one.

The base case of the recursion, in \mbox{Figure~\ref{fig:rem_exp_algo} Line 9},
is a (possibly negated) atomic constraint.
\mbox{Lines 10--13} determine which operators are included in
the remediated pipeline.
If a constraint notes that an operator's hyperparameter must be
\textit{present} or cannot be \textit{absent} and the corresponding
operator is part of a choice, \python{restrictChoice} removes that
choice from the pipeline in favor of the required operator.

\begin{figure}
\begin{lstlisting}[language=python,frame=single,framerule=0.5pt,backgroundcolor=\color{background},xleftmargin=2pt, xrightmargin=2pt]
pca = PCA.customize_schema(n_components=features_schema)
select_k_best = SelectKBest.customize_schema(k=features_schema)
planned = pca >> select_k_best >> LogisticRegression
\end{lstlisting}
\vspace{-5mm}
\caption{Example pipeline (g).}
\label{fig:expl-g}
\vspace{-5mm}
\end{figure}

Line~14 detects whether the constraint involves multiple
hyperparameters (possibly from multiple operators), such as in
pipeline~(g) in Figure~\ref{fig:expl-g}, where \python{PCA.n_components}
must be less than~\python{SelectKBest.k} because otherwise too few columns would
be piped to \python{SelectKBest}.
In these cases, because schemas are modularized per-operator, and
because JSON schema cannot express a less-than constraint involving
two hyperparameters, function \python{makeComparison} in Line~15
proxies this constraint by splitting the
possible values for the non-dependent hyperparameter into a number of
ranges (our default is five).
For example, if \python{k} can range from \mbox{5..55}, then five
versions of the \python{SelectKBest} operator are created where
\python{k} may range from \mbox{5..15}, \mbox{16..25}, ..., \mbox{46..55}.
Then, the dependent hyperparameter is also split such that it complies
to the constraint.
For example, if \python{n_components} originally ranged from \mbox{1..40},
then five versions of \python{PCA} are created where \python{n_components}
may range from \mbox{1..4}, \mbox{1..15}, ..., \mbox{1..40}, thus
guaranteeing that it is less than the corresponding \python{k} range.
Finally, \python{makeComparison} combines these pairs via Lale's
choice combinator (\python{|}).

Lastly, Line~17 handles the simplest and most common case of applying
constraints to a single hyperparameter and operator.
Constraints may either limit a hyperparameter to a set of values or,
if negated, exclude them from a given set of values.
To apply such constraints, we use Lale's
\pyplain{customize\_schema} feature, which returns a copy of an operator
that specifies a different schema for one of its hyperparameters.
Recent work shows how to make JSON Schema closed under conjunction and
negation~\cite{baazizi_et_al_2020}, but since that work is not
open-source, we implemented our own.
We translate a given constraint into the corresponding schema, as in
the example in Section~\ref{sec:detailed_example} that restricts
the \pyplain{strategy} hyperparameter of the \pyplain{SimpleImputer}
operator to the value \pyplain{"most\_frequent"}.

\tool's remediator is flexible enough to be used with other localization
algorithms so long as they output constraints in a compatible
format.
We implemented alternative algorithms and successfully used them with
our remediator as part of our evaluation, as described in more detail
in Section~\ref{sec:correctness}.

\subsection{Explanation}\label{sec:explanation}

The final component of \tool is an explainer that assists the user in
understanding the suggested remediation found by the solver via natural language.
Similar to the remediator, \tool's explanation features are flexible
enough to be used with other localization methods as long as they output
constraints in a compatible format.

Creating a natural language explanation uses a similar algorithm as
that for remediation in Figure~\ref{fig:rem_exp_algo}.
The main difference is that the helper functions \python{makeChoice},
\python{restrictChoice}, \python{makeComparison}, and
\python{customizeSchema} generate natural language instead of Python code.
For instance, \python{makeChoice} for explanation simply joins
constraints using the English word ``OR'' and newlines.
The other difference is that Line~8, instead of making a chained
call on the output of the previous step, uses the English word ``and.''
For a full example, consider the explanation for example pipeline~(k):

\begin{lstlisting}[frame=single,framerule=0pt,aboveskip=1mm]
Try setting argument 'with_mean' in operator StandardScaler to 'False'
OR
Try setting argument 'with_mean' in operator StandardScaler to 'True'
 and try ensuring that argument 'handle_unknown' in operator OrdinalEncoder
 is present for all runs (a Choice operator may need to be removed)
\end{lstlisting}
\vspace{-5mm}

\section{Evaluation}\label{sec:evaluation}

This section presents experiments for three research questions:

\begin{description}[leftmargin=9.2mm]
  \item[RQ1:] How does \tool's remediation affect correctness compared to baseline approaches?
  \item[RQ2:] How does \tool's remediation affect accuracy?
  \item[RQ3:] Does \tool's remediation help converge to optimal configurations more quickly?
\end{description}

We include the pipelines, results, and version of \tool used in this evaluation as part of the replication kit\footnote{https://zenodo.org/record/6385800}.

%
%

\subsection{Baseline Localization Algorithms}\label{sec:baseline}

We compare the correctness of \tool to other baseline ML fault localization
algorithms: modified versions of the \textit{Shortcut} and
\textit{Stacked Shortcut} methods from BugDoc~\cite{lourenco2020}.
These algorithms only attempt to identify and report root causes for
failures as constraints and do not include remediation, so we convert
the reported root causes into a compatible format for \tool's
remediator and explainer.

\subsection{Correctness (RQ1)}\label{sec:correctness}


\tool is inherently a correctness tool: given a set of evaluations,
some of which are incorrect, it locates the fault and repairs the
planned pipeline.
However, when a new AutoML search is launched starting from the
remediated planned pipeline, it will almost certainly attempt new
pipeline instances that \tool has not seen before.
There is no a priori guarantee that those new instances do
not fail in new ways.

Our evaluation set is the set of 20 pipeline use-cases described in
Section~\ref{sec:examples} which cover a wide
variety of potential failure cases.
To create this set, we started with problematic planned pipelines mined from
OpenML, plus data scientist interviews mentioning common failure causes.
After that, one author created additional cases to challenge our tool, drawing
upon documented constraints, Python raise statements, and reported issues.
For each example pipeline, we report whether each method was
able to find a remediation and whether failures occured after 20 more
AutoML-generated evaluations based on the remediation.

\begin{table}
\centerline{\begin{tabular}{@{}l | r | r | r@{}}
Localization & Successful & Restrictive & Unsuccessful \\
\hline
\tool             &    17 &    5 &   3 \\
Shortcut          &     7 &    1 &  13 \\
Stacked Shortcut  &     7 &    2 &  13 \\
\end{tabular}}
\vspace*{2mm}
\caption{\label{tab:correctness}Correctness evaluation per localization method.}
\vspace{-10mm}
\end{table}

Table~\ref{tab:correctness} summarizes the results.
A remediation is considered successful if it generates
no failures after 20 more AutoML-generated evaluations based
on the remediation.
\tool is able to successfully remediate all but three cases, whereas
the baseline methods are only able to successfully determine root
causes in seven cases each.
(Four of these successful remediations are due to examples that
require removing operators from choices which are part of our
modifications.
Without such modifications, the number of successes would be lower.)
In five examples, \tool suggests a fix that is
more restrictive but does not generate failures.
A restrictive remediation is one that may
restrict the potential search space for an AutoML pipeline more than a
manual remediation.
This may be due to a limitation of this evaluation method where
\tool only has access to 20 automatically-generated examples which may
insufficiently cover the space of expected fixes.
For example, an ideal remediation for a pipeline may be a constraint where
\mbox{\python{n_neighbors}$\le$15}.
With an input of 20 evaluations, \tool suggested a constraint of
\mbox{$\le$8}, which is more restrictive but technically correct.
Increasing the input to 50 evaluations increased the constraint
to \mbox{$\le$13}, which is closer to the ideal remediation.
We expect that both the restrictive and unsuccessful remediations might be
improved with additional input evaluations or a second round of
remediation.
We note that because \tool supports a full round-trip, we are able to
perform successive automated debugging on unsuccessful remediated pipelines.

The baseline methods insufficiently find root causes for a number of
reasons.
One reason is that they are simply not expressive
enough to successfully remediate the pipeline.
One example is example pipeline~(k) as seen in Figure~\ref{fig:expl-k}, where
\python{with_mean} only has a constraint depending on
the encoder selected.
The baseline methods only express simple equality or inequality
constraints.
Simply reporting a single constraint or even a union of constraints is
insufficient to describe this remediation.
Another reason is that the baseline methods assume that
hyperparameters are independent and can be freely
swapped without additional consequences.
However, hyperparameters are sometimes dependent on each other even across
operators, such as in pipeline~(g) in Figure~\ref{fig:expl-g}, where \python{k}
must be \mbox{$\le$\python{n_components}}.
Lastly, the baseline methods each only
consider a single failing pipeline instance whereas \tool considers
all failing instances.
Although we did not implement the \textit{Debugging Decision Tree}
method from BugDoc~\cite{lourenco2020}, we expect that if we modified
it in similar ways to the \textit{Shortcut} method and augmented it
with our automated remediation, it would fail to find remediations for
many of the cases for similar reasons. It is also highly expensive
(exponential time) to run for a realistic ML pipeline
so we chose not to reimplement it, especially given that we expect
similar performance to \textit{Shortcut}.

\subsection{Accuracy (RQ2) and Convergence (RQ3)}\label{sec:accuracy}

Although \tool focuses on correctness, that must be balanced with
predictive performance on the given dataset.
Since \tool is designed to work with AutoML tools, it is possible that
the remediation may remove too much of the potential search space in
order to guarantee a correct pipeline.
We would then expect new AutoML searches on this remediated pipeline
to also perform poorly.
We compare the AutoML predictive performance of the original pipeline to that of
the remediated pipeline provided by \tool in terms of test set
accuracy and number of iterations.

For the accuracy evaluations, we run two AutoML jobs: the original
pipeline and the remediated pipeline returned by \tool after 20
evaluations of the first job.
We use a train$+$test split of 80\%$+$20\% for the given
dataset and run each AutoML job for 1,000 iterations for both the
original and remediated pipeline using Hyperopt~\cite{hyperopt}.
Let ``optimal'' accuracy refer to the best test set accuracy
discovered in 1,000 iterations~\cite{arcuri2014}.
We run each job five times and report the average optimal accuracy
discovered by the five identical AutoML jobs and the average
number of iterations taken to reach it.

Remediated pipelines
have better optimal accuracy in \mbox{8 out of 20} cases and the same
optimal accuracy in an additional \mbox{7 out of 20} cases, while the
original pipeline has better accuracy in the remaining five cases.
For the cases with differing accuracies, the average difference is
relatively small at 0.0049, and a paired t-test suggests that original
and remediated optimal accuracies do not vary significantly (p=0.149).
For RQ2, this suggests that the remediations created by \tool on
average do not reduce accuracy and therefore are not removing
potentially beneficial sections of the search space.

We also examine the number of iterations to reach optimal accuracy,
specifically for the 15 cases where the remediated
pipeline discovers a better or equal optimal accuracy than the original.
We focus on these cases to have a similar point of comparison in terms
of iterations needed. We compare the average number of iterations needed for
the remediated pipeline to match or surpass the average optimal accuracy of
the original. In these cases, more remediated pipelines reach
the original pipeline's optimum accuracy faster (10 out of 15). The original
pipeline is faster in four cases while in one case both reach the
optimal accuracy in the same average number of iterations.
However, a paired t-test suggests that the average iterations for original and remediated pipelines to discover the original optimal accuracy do not vary significantly (p=0.935).
For RQ3, this suggests that remediations created by \tool on average also do
not change time to convergence compared to original pipelines.



%


\section{Threats to Validity}\label{sec:discussion}

The biggest limitation might appear to be that we only show
remediations based on the five formulae $S_1$~to $S_5$ in
Section~\ref{sec:system}.
However, these formulae turn out to be sufficient for all
20 planned pipelines used in the evaluation.
These formulae are much more expressive than similar tools
which are unable to cover all example pipelines.
We also note that these formulae could easily be extended for \tool.
Another limitation is that our tool does not guarantee finding minimal
root causes for all possible instances of a given planned pipeline but only
finds a root cause for a set of given instances, usually generated by an
AutoML job. However, our experiments (RQ1) suggest that a relatively modest
number of instances (20) is enough correctly remediate a pipeline in most
cases.
In the evaluation of \tool, we did not perform a human user study to examine
usability. Although we do consider usability valuable, our central claims of
correctness, accuracy, and convergence do not rely on human studies for
evaluation.


\section{Related Work}

\paragraph{AI development and AutoML} Systems with AI components come with unique challenges for the engineering process and individual developers. Debugging in particular is challenging due to errors hiding in data rather than code~\cite{Wan2019, humbatova2020} and the sheer amount of effort involved and the impossibility of manual evaluation due to the potentially millions of parameters to inspect~\cite{hill2016, arpteg2018, islam2019}.
To reduce this burden of manually exploring ML models, automated machine learning (AutoML) tools such as auto-sklearn~\cite{feurer2015} and AutoKeras~\cite{jin2019} use Bayesian optimizers to automatically construct ML pipelines and their hyperparameters.
We position our work among engineering tools that expand the capabilities of AutoML rather than improve the search or optimization performance.
One such tool is Lale~\cite{baudart_et_al_2020,baudart_et_al_2021}, which is a library of Python interfaces around ML operators designed to provide a consistent method of specifying pipelines for AutoML. \tool takes advantage of Lale's ability to precisely specify the search space for the pipeline's operators and hyperparameters.
Another such tool is AMS~\cite{cambronero2020}, which automatically ``strengthens'' weak pipeline specifications for AutoML by providing alternative operators and suggested hyperparameter spaces to search via learning over an existing corpus of AI software.

\paragraph{Data and AI debugging tools} We position our work among other tools
that assist developers in debugging and troubleshooting data-centric and AI
software.
Data-centric tools such as Panda~\cite{ikeda2012} and
PerfDebug~\cite{perfdebug2019} use data provenance to aid in debugging
data-centric pipelines and post-mortem performance issues respectively.
BigSift~\cite{bigsift2018} automatically generates a minimum set of inputs that
reproduce a test failure when given an Apache Spark program, test oracle, and
input dataset.
Dagger~\cite{dagger2020} is an end-to-end system for debugging data-centric
errors in pipelines where users manually instrument their Python code for later
logging and querying.
AI debugging tools such as LEMON~\cite{wang2020} and Haq et al.~\cite{haq2021}
automatically generate test suites for deep learning frameworks and KP-DNNs
respectively.
Nushi et al.~\cite{nushi2017} describe a human-in-the-loop methodology for
troubleshooting AI systems which uses crowd-sourcing to simulate potential
fixes to components.
DialTest~\cite{dialtest2021} is a tool for automatically detecting faults in
RNN-driven dialogue systems using transformations guided by Gini impurity.
Habib et al.\ use JSON subschema checks to find bugs in
ML pipelines~\cite{habib_et_al_2021}.
Though not AI-related, our tool is also related to constraint-based automated
program repair tools such as SemFix~\cite{ngyuen2013}.

The tool closest to our work is BugDoc~\cite{lourenco2020}, which automatically infers root causes of failures in ML pipelines based on previous executions. This is similar in concept to \tool's localizer component. BugDoc does not attempt to remediate nor further explain root causes unlike \tool. To our knowledge, \tool is the first tool to implement automated remediation and natural language and visual constraint explanation components in the context of debugging ML pipelines. Our tool's localizer component also differs from BugDoc in that it is designed for planned pipelines that work with AutoML and is also more expressive. Debugging support for such pipelines is more complex than normal ML pipelines in that each pipeline is a search space where operators and parameters may vary. To our knowledge, \tool is the first tool of its kind to express complex constraints between hyperparameters, operators, and other constraints. This is reflected in the evaluation in Section~\ref{sec:correctness} where baselines based on BugDoc algorithms are unable to identify root causes in most cases whereas \tool is able to. The aforementioned section details the differences between BugDoc and \tool's localizer component as they relate to our empirical evaluation.

We note that among all of the related automated (i.e.~\cite{bigsift2018, lourenco2020, haq2021, dialtest2021}) or end-to-end (i.e.~\cite{dagger2020}) debugging tools, \tool is unique in that it not only automatically identifies bugs but remediates them as well rather than only identifying root causes or generating test cases. This round-trip from actionable ML pipeline to remediated actionable ML pipeline is, to our knowledge, unique to \tool.

\section{Broader Impacts}

Automated machine learning (AutoML) in general encourages computationally-heavy
approaches to common data science tasks which raise CO\textsubscript{2} emissions. We believe that our tool encourages less computational waste by
enabling data scientists to more efficiently use AutoML by not wasting
resources on failing combinations of operators and hyperparameters. We also
hope that \tool enables data scientists to fix errors faster and run less
AutoML jobs overall. However, it is also possible that helping data scientists
more easily debug AutoML pipelines may encourage further usage of automated
techiniques which may overall raise CO\textsubscript{2} emissions. Future work
may explore techniques to reduce the amount of initial AutoML iterations
necessary to remediate pipelines in order to encourage less wasteful automated
machine learning.

\bibliographystyle{ACM-Reference-Format}
\bibliography{main}


\begin{thebibliography}{31}


\ifx \showCODEN    \undefined \def \showCODEN     #1{\unskip}     \fi
\ifx \showDOI      \undefined \def \showDOI       #1{#1}\fi
\ifx \showISBNx    \undefined \def \showISBNx     #1{\unskip}     \fi
\ifx \showISBNxiii \undefined \def \showISBNxiii  #1{\unskip}     \fi
\ifx \showISSN     \undefined \def \showISSN      #1{\unskip}     \fi
\ifx \showLCCN     \undefined \def \showLCCN      #1{\unskip}     \fi
\ifx \shownote     \undefined \def \shownote      #1{#1}          \fi
\ifx \showarticletitle \undefined \def \showarticletitle #1{#1}   \fi
\ifx \showURL      \undefined \def \showURL       {\relax}        \fi
\providecommand\bibfield[2]{#2}
\providecommand\bibinfo[2]{#2}
\providecommand\natexlab[1]{#1}
\providecommand\showeprint[2][]{arXiv:#2}

\bibitem[Amershi et~al\mbox{.}(2019)]%
        {amershi2019}
\bibfield{author}{\bibinfo{person}{Saleema Amershi}, \bibinfo{person}{Andrew
  Begel}, \bibinfo{person}{Christian Bird}, \bibinfo{person}{Robert DeLine},
  \bibinfo{person}{Harald Gall}, \bibinfo{person}{Ece Kamar},
  \bibinfo{person}{Nachiappan Nagappan}, \bibinfo{person}{Besmira Nushi}, {and}
  \bibinfo{person}{Thomas Zimmermann}.} \bibinfo{year}{2019}\natexlab{}.
\newblock \showarticletitle{{Software Engineering for Machine Learning: A Case
  Study}}. In \bibinfo{booktitle}{\emph{International Conference on Software
  Engineering: Software Engineering in Practice (ICSE-SEIP)}}.
  \bibinfo{pages}{291--300}.
\newblock
\urldef\tempurl%
\url{https://doi.org/10.1109/ICSE-SEIP.2019.00042}
\showURL{%
\tempurl}


\bibitem[Arcuri and Briand(2014)]%
        {arcuri2014}
\bibfield{author}{\bibinfo{person}{Andrea Arcuri} {and} \bibinfo{person}{Lionel
  Briand}.} \bibinfo{year}{2014}\natexlab{}.
\newblock \showarticletitle{{A Hitchhiker's guide to statistical tests for
  assessing randomized algorithms in software engineering}}.
\newblock \bibinfo{journal}{\emph{Software Testing, Verification and
  Reliability}} \bibinfo{volume}{24}, \bibinfo{number}{3}
  (\bibinfo{year}{2014}), \bibinfo{pages}{219--250}.
\newblock
\urldef\tempurl%
\url{https://doi.org/10.1002/stvr.1486}
\showDOI{\tempurl}


\bibitem[Arpteg et~al\mbox{.}(2018)]%
        {arpteg2018}
\bibfield{author}{\bibinfo{person}{A Arpteg}, \bibinfo{person}{B Brinne},
  \bibinfo{person}{L Crnkovic-Friis}, {and} \bibinfo{person}{J Bosch}.}
  \bibinfo{year}{2018}\natexlab{}.
\newblock \showarticletitle{{Software Engineering Challenges of Deep
  Learning}}. In \bibinfo{booktitle}{\emph{Conference on Software Engineering
  and Advanced Applications (SEAA)}}. \bibinfo{pages}{50--59}.
\newblock
\urldef\tempurl%
\url{https://doi.org/10.1109/SEAA.2018.00018}
\showDOI{\tempurl}


\bibitem[Baazizi et~al\mbox{.}(2020)]%
        {baazizi_et_al_2020}
\bibfield{author}{\bibinfo{person}{Mohamed-Amine Baazizi},
  \bibinfo{person}{Dario Colazzo}, \bibinfo{person}{Giorgio Ghelli},
  \bibinfo{person}{Carlo Sartiani}, {and} \bibinfo{person}{Stefanie
  Scherzinger}.} \bibinfo{year}{2020}\natexlab{}.
\newblock \showarticletitle{Not Elimination and Witness Generation for {JSON}
  {Schema}}. In \bibinfo{booktitle}{\emph{Conf{\'e}rence sur la Gestion de
  Donn{\'e}es (BDA)}}.
\newblock
\urldef\tempurl%
\url{https://arxiv.org/abs/2104.14828}
\showURL{%
\tempurl}


\bibitem[Baudart et~al\mbox{.}(2020)]%
        {baudart_et_al_2020}
\bibfield{author}{\bibinfo{person}{Guillaume Baudart}, \bibinfo{person}{Martin
  Hirzel}, \bibinfo{person}{Kiran Kate}, \bibinfo{person}{Parikshit Ram}, {and}
  \bibinfo{person}{Avraham Shinnar}.} \bibinfo{year}{2020}\natexlab{}.
\newblock \showarticletitle{Lale: Consistent Automated Machine Learning}. In
  \bibinfo{booktitle}{\emph{KDD Workshop on Automation in Machine Learning
  (AutoML@KDD)}}.
\newblock
\urldef\tempurl%
\url{https://arxiv.org/abs/2007.01977}
\showURL{%
\tempurl}


\bibitem[Baudart et~al\mbox{.}(2021)]%
        {baudart_et_al_2021}
\bibfield{author}{\bibinfo{person}{Guillaume Baudart}, \bibinfo{person}{Martin
  Hirzel}, \bibinfo{person}{Kiran Kate}, \bibinfo{person}{Parikshit Ram},
  \bibinfo{person}{Avraham Shinnar}, {and} \bibinfo{person}{Jason Tsay}.}
  \bibinfo{year}{2021}\natexlab{}.
\newblock \showarticletitle{Pipeline Combinators for Gradual {AutoML}}. In
  \bibinfo{booktitle}{\emph{Advances in Neural Information Processing Systems
  (NeurIPS)}}.
\newblock
\urldef\tempurl%
\url{https://proceedings.neurips.cc/paper/2021/file/a3b36cb25e2e0b93b5f334ffb4e4064e-Paper.pdf}
\showURL{%
\tempurl}


\bibitem[Bellamy et~al\mbox{.}(2019)]%
        {bellamy2019}
\bibfield{author}{\bibinfo{person}{R~K~E Bellamy}, \bibinfo{person}{K Dey},
  \bibinfo{person}{M Hind}, \bibinfo{person}{S~C Hoffman}, \bibinfo{person}{S
  Houde}, \bibinfo{person}{K Kannan}, \bibinfo{person}{P Lohia},
  \bibinfo{person}{J Martino}, \bibinfo{person}{S Mehta}, \bibinfo{person}{A
  Mojsilovi{\'{c}}}, \bibinfo{person}{S Nagar}, \bibinfo{person}{K~N
  Ramamurthy}, \bibinfo{person}{J Richards}, \bibinfo{person}{D Saha},
  \bibinfo{person}{P Sattigeri}, \bibinfo{person}{M Singh},
  \bibinfo{person}{K~R Varshney}, {and} \bibinfo{person}{Y Zhang}.}
  \bibinfo{year}{2019}\natexlab{}.
\newblock \showarticletitle{{AI Fairness 360: An extensible toolkit for
  detecting and mitigating algorithmic bias}}.
\newblock \bibinfo{journal}{\emph{IBM Journal of Research and Development}}
  \bibinfo{volume}{63}, \bibinfo{number}{4/5} (\bibinfo{date}{jul}
  \bibinfo{year}{2019}), \bibinfo{pages}{4:1--4:15}.
\newblock
\showISSN{0018-8646}
\urldef\tempurl%
\url{https://doi.org/10.1147/JRD.2019.2942287}
\showDOI{\tempurl}


\bibitem[Bergstra et~al\mbox{.}(2013)]%
        {hyperopt}
\bibfield{author}{\bibinfo{person}{James Bergstra}, \bibinfo{person}{Daniel
  Yamins}, {and} \bibinfo{person}{David Cox}.} \bibinfo{year}{2013}\natexlab{}.
\newblock \showarticletitle{Making a Science of Model Search: Hyperparameter
  Optimization in Hundreds of Dimensions for Vision Architectures}. In
  \bibinfo{booktitle}{\emph{International Conference on Machine Learning
  (ICML)}}. \bibinfo{pages}{115--123}.
\newblock


\bibitem[Cambronero et~al\mbox{.}(2020)]%
        {cambronero2020}
\bibfield{author}{\bibinfo{person}{Jos\'{e}~P. Cambronero},
  \bibinfo{person}{J\"{u}rgen Cito}, {and} \bibinfo{person}{Martin~C. Rinard}.}
  \bibinfo{year}{2020}\natexlab{}.
\newblock \showarticletitle{AMS: Generating AutoML Search Spaces from Weak
  Specifications}. In \bibinfo{booktitle}{\emph{Joint Meeting on European
  Software Engineering Conference and Symposium on the Foundations of Software
  Engineering (ESEC/FSE)}}. \bibinfo{pages}{763--774}.
\newblock
\urldef\tempurl%
\url{https://doi.org/10.1145/3368089.3409700}
\showURL{%
\tempurl}


\bibitem[Feurer et~al\mbox{.}(2015)]%
        {feurer2015}
\bibfield{author}{\bibinfo{person}{Matthias Feurer}, \bibinfo{person}{Aaron
  Klein}, \bibinfo{person}{Katharina Eggensperger}, \bibinfo{person}{Jost
  Springenberg}, \bibinfo{person}{Manuel Blum}, {and} \bibinfo{person}{Frank
  Hutter}.} \bibinfo{year}{2015}\natexlab{}.
\newblock \showarticletitle{Efficient and Robust Automated Machine Learning}.
\newblock In \bibinfo{booktitle}{\emph{Conference on Neural Information
  Processing Systems (NIPS)}}. \bibinfo{pages}{2962--2970}.
\newblock
\urldef\tempurl%
\url{http://papers.nips.cc/paper/5872-efficient-and-robust-automated-machine-learning.pdf}
\showURL{%
\tempurl}


\bibitem[Gulzar et~al\mbox{.}(2018)]%
        {bigsift2018}
\bibfield{author}{\bibinfo{person}{Muhammad~Ali Gulzar}, \bibinfo{person}{Siman
  Wang}, {and} \bibinfo{person}{Miryung Kim}.} \bibinfo{year}{2018}\natexlab{}.
\newblock \showarticletitle{{BigSift}: Automated Debugging of Big Data
  Analytics in Data-Intensive Scalable Computing}. In
  \bibinfo{booktitle}{\emph{Joint Meeting on European Software Engineering
  Conference and Symposium on the Foundations of Software Engineering
  (ESEC/FSE)}}. \bibinfo{pages}{863--866}.
\newblock
\urldef\tempurl%
\url{https://doi.org/10.1145/3236024.3264586}
\showURL{%
\tempurl}


\bibitem[Gundersen and Kjensmo(2017)]%
        {gundersen2017}
\bibfield{author}{\bibinfo{person}{Odd~Erik Gundersen} {and}
  \bibinfo{person}{Sigbj{\o}rn Kjensmo}.} \bibinfo{year}{2017}\natexlab{}.
\newblock \showarticletitle{State of the Art: Reproducibility in Artificial
  Intelligence}. In \bibinfo{booktitle}{\emph{Conference on Artificial
  Intelligence (AAAI)}}. \bibinfo{pages}{1644--1651}.
\newblock
\urldef\tempurl%
\url{https://ojs.aaai.org/index.php/AAAI/article/view/11503}
\showURL{%
\tempurl}


\bibitem[Habib et~al\mbox{.}(2021)]%
        {habib_et_al_2021}
\bibfield{author}{\bibinfo{person}{Andrew Habib}, \bibinfo{person}{Avraham
  Shinnar}, \bibinfo{person}{Martin Hirzel}, {and} \bibinfo{person}{Michael
  Pradel}.} \bibinfo{year}{2021}\natexlab{}.
\newblock \showarticletitle{Finding Data Compatibility Bugs with {JSON}
  Subschema Checking}. In \bibinfo{booktitle}{\emph{International Symposium on
  Software Testing and Analysis (ISSTA)}}. \bibinfo{pages}{620--632}.
\newblock
\urldef\tempurl%
\url{https://doi.org/10.1145/3460319.3464796}
\showURL{%
\tempurl}


\bibitem[Haq et~al\mbox{.}(2021)]%
        {haq2021}
\bibfield{author}{\bibinfo{person}{Fitash~Ul Haq}, \bibinfo{person}{Donghwan
  Shin}, \bibinfo{person}{Lionel~C Briand}, \bibinfo{person}{Thomas Stifter},
  {and} \bibinfo{person}{Jun Wang}.} \bibinfo{year}{2021}\natexlab{}.
\newblock \showarticletitle{{Automatic Test Suite Generation for Key-Points
  Detection DNNs Using Many-Objective Search (Experience Paper)}}. In
  \bibinfo{booktitle}{\emph{International Symposium on Software Testing and
  Analysis (ISSTA)}}. \bibinfo{publisher}{Association for Computing Machinery},
  \bibinfo{pages}{91--102}.
\newblock
\showISBNx{9781450384599}
\urldef\tempurl%
\url{https://doi.org/10.1145/3460319.3464802}
\showDOI{\tempurl}


\bibitem[Hill et~al\mbox{.}(2016)]%
        {hill2016}
\bibfield{author}{\bibinfo{person}{C Hill}, \bibinfo{person}{R Bellamy},
  \bibinfo{person}{T Erickson}, {and} \bibinfo{person}{M Burnett}.}
  \bibinfo{year}{2016}\natexlab{}.
\newblock \showarticletitle{{Trials and tribulations of developers of
  intelligent systems: A field study}}. In \bibinfo{booktitle}{\emph{Symposium
  on Visual Languages and Human-Centric Computing (VL/HCC)}}.
  \bibinfo{pages}{162--170}.
\newblock
\urldef\tempurl%
\url{https://doi.org/10.1109/VLHCC.2016.7739680}
\showDOI{\tempurl}


\bibitem[Humbatova et~al\mbox{.}(2020)]%
        {humbatova2020}
\bibfield{author}{\bibinfo{person}{Nargiz Humbatova}, \bibinfo{person}{Gunel
  Jahangirova}, \bibinfo{person}{Gabriele Bavota}, \bibinfo{person}{Vincenzo
  Riccio}, \bibinfo{person}{Andrea Stocco}, {and} \bibinfo{person}{Paolo
  Tonella}.} \bibinfo{year}{2020}\natexlab{}.
\newblock \showarticletitle{Taxonomy of Real Faults in Deep Learning Systems}.
  In \bibinfo{booktitle}{\emph{International Conference on Software Engineering
  (ICSE)}}. \bibinfo{pages}{1110--1121}.
\newblock
\urldef\tempurl%
\url{https://doi.org/10.1145/3377811.3380395}
\showDOI{\tempurl}


\bibitem[Ikeda et~al\mbox{.}(2012)]%
        {ikeda2012}
\bibfield{author}{\bibinfo{person}{Robert Ikeda}, \bibinfo{person}{Junsang
  Cho}, \bibinfo{person}{Charlie Fang}, \bibinfo{person}{Semih Salihoglu},
  \bibinfo{person}{Satoshi Torikai}, {and} \bibinfo{person}{Jennifer Widom}.}
  \bibinfo{year}{2012}\natexlab{}.
\newblock \showarticletitle{Provenance-Based Debugging and Drill-Down in
  Data-Oriented Workflows}. In \bibinfo{booktitle}{\emph{International
  Conference on Data Engineering (ICDE)}}. \bibinfo{pages}{1249--1252}.
\newblock
\urldef\tempurl%
\url{https://doi.org/10.1109/ICDE.2012.118}
\showURL{%
\tempurl}


\bibitem[Islam et~al\mbox{.}(2019)]%
        {islam2019}
\bibfield{author}{\bibinfo{person}{Md~Johirul Islam}, \bibinfo{person}{Giang
  Nguyen}, \bibinfo{person}{Rangeet Pan}, {and} \bibinfo{person}{Hridesh
  Rajan}.} \bibinfo{year}{2019}\natexlab{}.
\newblock \showarticletitle{{A Comprehensive Study on Deep Learning Bug
  Characteristics}}. In \bibinfo{booktitle}{\emph{Joint Meeting on European
  Software Engineering Conference and Symposium on the Foundations of Software
  Engineering (ESEC/FSE)}}. \bibinfo{pages}{510--520}.
\newblock
\urldef\tempurl%
\url{https://doi.org/10.1145/3338906.3338955}
\showURL{%
\tempurl}


\bibitem[Jin et~al\mbox{.}(2019)]%
        {jin2019}
\bibfield{author}{\bibinfo{person}{Haifeng Jin}, \bibinfo{person}{Qingquan
  Song}, {and} \bibinfo{person}{Xia Hu}.} \bibinfo{year}{2019}\natexlab{}.
\newblock \showarticletitle{{Auto-Keras}: An Efficient Neural Architecture
  Search System}. In \bibinfo{booktitle}{\emph{Conference on Knowledge
  Discovery and Data Mining (KDD)}}. \bibinfo{pages}{1946--1956}.
\newblock
\urldef\tempurl%
\url{http://doi.acm.org/10.1145/3292500.3330648}
\showURL{%
\tempurl}


\bibitem[Ke et~al\mbox{.}(2017)]%
        {ke_et_al_2017}
\bibfield{author}{\bibinfo{person}{Guolin Ke}, \bibinfo{person}{Qi Meng},
  \bibinfo{person}{Thomas Finley}, \bibinfo{person}{Taifeng Wang},
  \bibinfo{person}{Wei Chen}, \bibinfo{person}{Weidong Ma},
  \bibinfo{person}{Qiwei Ye}, {and} \bibinfo{person}{Tie-Yan Liu}.}
  \bibinfo{year}{2017}\natexlab{}.
\newblock \showarticletitle{{LightGBM}: A Highly Efficient Gradient Boosting
  Decision Tree}. In \bibinfo{booktitle}{\emph{Conference on Neural Information
  Processing Systems (NIPS)}}. \bibinfo{pages}{3146--3154}.
\newblock
\urldef\tempurl%
\url{http://papers.nips.cc/paper/6907-lightgbm-a-highly-efficient-gradient-boosting-decision-tree}
\showURL{%
\tempurl}


\bibitem[Liu et~al\mbox{.}(2021)]%
        {dialtest2021}
\bibfield{author}{\bibinfo{person}{Zixi Liu}, \bibinfo{person}{Yang Feng},
  {and} \bibinfo{person}{Zhenyu Chen}.} \bibinfo{year}{2021}\natexlab{}.
\newblock \showarticletitle{{DialTest}: Automated Testing for
  Recurrent-Neural-Network-Driven Dialogue Systems}. In
  \bibinfo{booktitle}{\emph{International Symposium on Software Testing and
  Analysis (ISSTA)}}. \bibinfo{pages}{115--126}.
\newblock
\urldef\tempurl%
\url{https://doi.org/10.1145/3460319.3464829}
\showURL{%
\tempurl}


\bibitem[Louren\c{c}o et~al\mbox{.}(2020)]%
        {lourenco2020}
\bibfield{author}{\bibinfo{person}{Raoni Louren\c{c}o},
  \bibinfo{person}{Juliana Freire}, {and} \bibinfo{person}{Dennis Shasha}.}
  \bibinfo{year}{2020}\natexlab{}.
\newblock \showarticletitle{BugDoc: A System for Debugging Computational
  Pipelines}. In \bibinfo{booktitle}{\emph{International Conference on
  Management of Data (SIGMOD)}}. \bibinfo{pages}{2733--2736}.
\newblock
\showISBNx{9781450367356}
\urldef\tempurl%
\url{https://doi.org/10.1145/3318464.3384692}
\showDOI{\tempurl}


\bibitem[{Nguyen} et~al\mbox{.}(2013)]%
        {ngyuen2013}
\bibfield{author}{\bibinfo{person}{H.~D.~T. {Nguyen}}, \bibinfo{person}{D.
  {Qi}}, \bibinfo{person}{A. {Roychoudhury}}, {and} \bibinfo{person}{S.
  {Chandra}}.} \bibinfo{year}{2013}\natexlab{}.
\newblock \showarticletitle{{SemFix}: Program Repair via Semantic Analysis}. In
  \bibinfo{booktitle}{\emph{International Conference on Software Engineering
  (ICSE)}}. \bibinfo{pages}{772--781}.
\newblock
\urldef\tempurl%
\url{https://doi.org/10.1109/ICSE.2013.6606623}
\showURL{%
\tempurl}


\bibitem[Nushi et~al\mbox{.}(2017)]%
        {nushi2017}
\bibfield{author}{\bibinfo{person}{Besmira Nushi}, \bibinfo{person}{Ece Kamar},
  \bibinfo{person}{Eric Horvitz}, {and} \bibinfo{person}{Donald Kossmann}.}
  \bibinfo{year}{2017}\natexlab{}.
\newblock \showarticletitle{On Human Intellect and Machine Failures:
  Troubleshooting Integrative Machine Learning Systems}. In
  \bibinfo{booktitle}{\emph{Conference on Artificial Intelligence (AAAI)}}.
  \bibinfo{pages}{1017--1025}.
\newblock
\urldef\tempurl%
\url{https://www.aaai.org/ocs/index.php/AAAI/AAAI17/paper/view/15032/0}
\showURL{%
\tempurl}


\bibitem[Pedregosa et~al\mbox{.}(2011)]%
        {scikit-learn}
\bibfield{author}{\bibinfo{person}{F. Pedregosa}, \bibinfo{person}{G.
  Varoquaux}, \bibinfo{person}{A. Gramfort}, \bibinfo{person}{V. Michel},
  \bibinfo{person}{B. Thirion}, \bibinfo{person}{O. Grisel},
  \bibinfo{person}{M. Blondel}, \bibinfo{person}{P. Prettenhofer},
  \bibinfo{person}{R. Weiss}, \bibinfo{person}{V. Dubourg}, \bibinfo{person}{J.
  Vanderplas}, \bibinfo{person}{A. Passos}, \bibinfo{person}{D. Cournapeau},
  \bibinfo{person}{M. Brucher}, \bibinfo{person}{M. Perrot}, {and}
  \bibinfo{person}{E. Duchesnay}.} \bibinfo{year}{2011}\natexlab{}.
\newblock \showarticletitle{Scikit-learn: Machine Learning in {P}ython}.
\newblock \bibinfo{journal}{\emph{Journal of Machine Learning Research (JMLR)}}
   \bibinfo{volume}{12} (\bibinfo{year}{2011}), \bibinfo{pages}{2825--2830}.
\newblock


\bibitem[Rezig et~al\mbox{.}(2020)]%
        {dagger2020}
\bibfield{author}{\bibinfo{person}{El~Kindi Rezig}, \bibinfo{person}{Ashrita
  Brahmaroutu}, \bibinfo{person}{Nesime Tatbul}, \bibinfo{person}{Mourad
  Ouzzani}, \bibinfo{person}{Nan Tang}, \bibinfo{person}{Timothy Mattson},
  \bibinfo{person}{Samuel Madden}, {and} \bibinfo{person}{Michael
  Stonebraker}.} \bibinfo{year}{2020}\natexlab{}.
\newblock \showarticletitle{Debugging Large-Scale Data Science Pipelines Using
  Dagger}. In \bibinfo{booktitle}{\emph{Demonstration at the Conference on Very
  Large Data Bases (VLDB-Demo)}}. \bibinfo{pages}{2993--2996}.
\newblock
\urldef\tempurl%
\url{https://doi.org/10.14778/3415478.3415527}
\showURL{%
\tempurl}


\bibitem[Sculley et~al\mbox{.}(2015)]%
        {sculley2015}
\bibfield{author}{\bibinfo{person}{D. Sculley}, \bibinfo{person}{Gary Holt},
  \bibinfo{person}{Daniel Golovin}, \bibinfo{person}{Eugene Davydov},
  \bibinfo{person}{Todd Phillips}, \bibinfo{person}{Dietmar Ebner},
  \bibinfo{person}{Vinay Chaudhary}, \bibinfo{person}{Michael Young},
  \bibinfo{person}{Jean-Fran\c{c}ois Crespo}, {and} \bibinfo{person}{Dan
  Dennison}.} \bibinfo{year}{2015}\natexlab{}.
\newblock \showarticletitle{Hidden Technical Debt in Machine Learning Systems}.
\newblock In \bibinfo{booktitle}{\emph{Conference on Neural Information
  Processing Systems (NIPS)}}. \bibinfo{pages}{2503--2511}.
\newblock
\urldef\tempurl%
\url{http://papers.nips.cc/paper/5656-hidden-technical-debt-in-machine-learning-systems}
\showURL{%
\tempurl}


\bibitem[Teoh et~al\mbox{.}(2019)]%
        {perfdebug2019}
\bibfield{author}{\bibinfo{person}{Jason Teoh}, \bibinfo{person}{Muhammad~Ali
  Gulzar}, \bibinfo{person}{Guoqing~Harry Xu}, {and} \bibinfo{person}{Miryung
  Kim}.} \bibinfo{year}{2019}\natexlab{}.
\newblock \showarticletitle{{PerfDebug}: Performance Debugging of Computation
  Skew in Dataflow Systems}. In \bibinfo{booktitle}{\emph{Symposium on Cloud
  Computing (SoCC)}}. \bibinfo{pages}{465--476}.
\newblock
\showISBNx{9781450369732}
\urldef\tempurl%
\url{https://doi.org/10.1145/3357223.3362727}
\showDOI{\tempurl}


\bibitem[Torlak and Bodik(2013)]%
        {torlak_bodik_2013}
\bibfield{author}{\bibinfo{person}{Emina Torlak} {and}
  \bibinfo{person}{Rastislav Bodik}.} \bibinfo{year}{2013}\natexlab{}.
\newblock \showarticletitle{Growing Solver-Aided Languages with Rosette}. In
  \bibinfo{booktitle}{\emph{Symposium on New Ideas, New Paradigms, and
  Reflections on Programming and Software (Onward!)}}.
  \bibinfo{pages}{135--152}.
\newblock
\urldef\tempurl%
\url{https://doi.org/10.1145/2509578.2509586}
\showURL{%
\tempurl}


\bibitem[Wan et~al\mbox{.}(2019)]%
        {Wan2019}
\bibfield{author}{\bibinfo{person}{Zhiyuan Wan}, \bibinfo{person}{Xin Xia},
  \bibinfo{person}{David Lo}, {and} \bibinfo{person}{Gail~C. Murphy}.}
  \bibinfo{year}{2019}\natexlab{}.
\newblock \showarticletitle{{How does Machine Learning Change Software
  Development Practices?}}
\newblock \bibinfo{journal}{\emph{Transactions on Software Engineering (TSE)}}
  (\bibinfo{year}{2019}).
\newblock
\urldef\tempurl%
\url{https://doi.org/10.1109/TSE.2019.2937083}
\showURL{%
\tempurl}


\bibitem[Wang et~al\mbox{.}(2020)]%
        {wang2020}
\bibfield{author}{\bibinfo{person}{Zan Wang}, \bibinfo{person}{Ming Yan},
  \bibinfo{person}{Junjie Chen}, \bibinfo{person}{Shuang Liu}, {and}
  \bibinfo{person}{Dongdi Zhang}.} \bibinfo{year}{2020}\natexlab{}.
\newblock \showarticletitle{Deep Learning Library Testing via Effective Model
  Generation}. In \bibinfo{booktitle}{\emph{Joint Meeting on European Software
  Engineering Conference and Symposium on the Foundations of Software
  Engineering (ESEC/FSE)}} \emph{(\bibinfo{series}{ESEC/FSE 2020})}.
  \bibinfo{pages}{788--799}.
\newblock
\showISBNx{9781450370431}
\urldef\tempurl%
\url{https://doi.org/10.1145/3368089.3409761}
\showDOI{\tempurl}


\end{thebibliography}

\appendix
\section{Appendix}

\subsection{Fault Localization}

 This appendix defines all the constraints more formally, following
 the discussion in Section~\ref{sec:localization}.  As discussed in that
 section, we use symbolic
 variables to write the following:$$S_1\left(\mathfrak{P}\right) \equiv \forall_{p \in \mathfrak{P}} \left( r(p)
    \Leftrightarrow \left<@\mathit{hparam}, @\mathit{value}\right> \in
  H(p)\right)$$This symbolic program is run by the solver as
$$\mathit{solve}\left(\mathit{assert}\left(S_1\left(\mathfrak{P}\right)\right)\right)$$
(the
  symbolic machinery does not change, so we hereafter omit the
  explicit $\mathit{solve}$ call) and returns a model, if one exists, that
binds the symbolic variables $@\mathit{hparam}$ and $@\mathit{value}$ to concrete values
that make the assertion true.  This will find any hyperparameter
setting that correlates exactly with pipelines that fail.  In fact,
this simple logic suffices for any failure caused by a single value of
a single hyperparameter.  Based  on this formalism, we present the
details that were omitted on Section~\ref{sec:localization}.

 A special case is when the mere presence of the hyperparameter is enough
to cause failure, meaning that that pipeline step must be removed.
Failed runs have this hyperparameter and successful ones do not, meaning
they do not have the step at all.  This can be expressed by quantifying over the
value:$$S_2 \equiv \forall_{p \in \mathfrak{P}} \left(
      \neg r(p)
    \Leftrightarrow \left<@hyper, \exists v . v\right> \in
  H(p)\right)$$

But problems can get more complex.  Consider pipeline~(e): the issue
is that \python{KNeighborsClassifier} needs \python{n_neighbors} to be
smaller than the size of the dataset, and this example was run on a
tiny dataset.  In this case, any value $\le16$ (the size of the
cross-validation folds of the training split) works, so we need a
constraint like:
$$S_3 \equiv \forall_{p \in \mathfrak{P}} \left( r(p)
      \Leftrightarrow \left(
         \begin{array}{l}
           \left<@\mathit{hparam}, @\mathit{value}\right> \in H(p) \;\wedge \\
           @\mathit{value} \; @op \; @limit
         \end{array} \right)\right)$$This will find
the requirement \mbox{\python{n_neighbors}$\,\le 16$}.  Rosette
supports both integer and real numbers and the usual operators on
them, so this supports any numerical limit on a single hyperparameter.

But some problems involve constraints, typically numerical ones,
between multiple hyperparameters.  Take pipeline (g),
where principal component analysis (\python{PCA})
reduces the number of features and then \python{SelectKBest}
chooses the best $k$ features.  To choose the $k$ best, there must be at
least $k$, so \python{PCA} must preserve enough.  This requires a constraint that
relates two hyperparameters:$$S_4 \equiv \forall_{p \in \mathfrak{P}} \left( r(p)
      \Leftrightarrow \left(
         \begin{array}{l}
           \left<@\mathit{hparam}_1, @\mathit{value}_1\right> \in H(p) \;\wedge \\
           \left<@\mathit{hparam}_2, @\mathit{value}_2\right> \in H(p) \;\wedge \\
           @\mathit{value}_1 \; @op \; @\mathit{value}_2
         \end{array} \right)\right)$$This will find that all
 successful pipeline instances from this example have
 \mbox{\python{PCA.n_components}$\,\leq\,$\python{SelectKBest.k}}.

 It is simple to make a general solver as
 follows:$$S_\mathit{any}\left(\mathfrak{P}\right) \equiv S_1(\mathfrak{P}) \vee S_2(\mathfrak{P}) \vee S_3(\mathfrak{P}) \vee S_4(\mathfrak{P})$$

 However, these constraints all assume that the same settings are
 needed across all pipelines, and that is not always the case.
 Consider pipeline~(f), in which the combination
\python{"whiten"} and the \python{"arpack"} solver breaks for this dataset.  Either is
allowed, but they cannot be used together.  To handle this, we stack
these constraints such that one constraint controls which other constraint
applies; the superscripts on $S$ indicate that the three uses of $S$
generate distinct symbolic variables, so there are three independent
constraints:$$
  S_5 \equiv \forall_{p \in \mathfrak{P}} \left( r(p)
    \Leftrightarrow
    \left\{
        \begin{array}{ll}
          S_\mathit{any}^2(\left\{p\right\}) & \textrm{if }S_\mathit{any}^1(\left\{p\right\}) \\
          S_\mathit{any}^3(\left\{p\right\}) & \textrm{otherwise}
        \end{array}\right.
\right)
$$
The values for all the constraints can be read directly from the
variables produced by the solver.  This only illustrates two levels, but clearly they can be
stacked as deeply as needed.

\subsection{List of Example Pipelines}

Table~\ref{fig:examples} contains the full set of pipelines used to evaluate
the system as described in Section~\ref{sec:examples}.

\begin{table*}
\begin{tabular}{@{}clll@{}}
& Planned pipeline & Cause of failure & Remediation
\\\hline
(a)
&
\begin{minipage}{1.025\columnwidth}
\begin{lstlisting}[language=python]
project = Project(columns={"type": "number"})
planned = (project | NoOp) >> LogisticRegression
\end{lstlisting}
\end{minipage}
&
\begin{minipage}{0.45\columnwidth}\raggedright\scriptsize
non-numbers for \python{LogisticRegression}
\end{minipage}
&
\begin{minipage}{0.45\columnwidth}\raggedright\scriptsize
remove \python{NoOp} from choice (\python{|})
to ensure projection to numbers
\end{minipage}
\\[-1mm]\hline\\[-4mm]
(b)
&
\begin{minipage}{1.025\columnwidth}
\begin{lstlisting}[language=python]
project = Project(columns={"type": "string"})
one_hot = OneHotEncoder(handle_unknown="ignore")
planned = project >> one_hot >> StandardScaler >> LogisticRegression
\end{lstlisting}
\end{minipage}
&
\begin{minipage}{0.45\columnwidth}\raggedright\scriptsize
\python{one_hot} returns sparse data and \python{StandardScaler} uses
(or defaults to) \python{with_mean=True}
\end{minipage}
&
\begin{minipage}{0.45\columnwidth}\raggedright\scriptsize
set \python{with_mean=False} in \python{StandardScaler}
\end{minipage}
\\[-1mm]\hline\\[-4mm]
(c)
&
\begin{minipage}{1.025\columnwidth}
\begin{lstlisting}[language=python]
project = Project(columns={"type": "string"})
one_hot = OneHotEncoder(handle_unknown="ignore")
planned = project >> SimpleImputer >> one_hot >> LogisticRegression
\end{lstlisting}
\end{minipage}
&
\begin{minipage}{0.45\columnwidth}\raggedright\scriptsize
\python{strategy in ["mean", "median"]}
only works for numeric missing values
\end{minipage}
&
\begin{minipage}{0.45\columnwidth}\raggedright\scriptsize
set \python{strategy="most_frequent"},
which also works for non-numeric values
\end{minipage}
\\[-1mm]\hline\\[-4mm]
(d)
&
\begin{minipage}{1.025\columnwidth}
\begin{lstlisting}[language=python]
planned = PCA >> GradientBoostingClassifier
\end{lstlisting}
\end{minipage}
&
\begin{minipage}{0.45\columnwidth}\raggedright\scriptsize
exponential loss, 3-class data
\end{minipage}
&
\begin{minipage}{0.45\columnwidth}\raggedright\scriptsize
set \python{loss="deviance"} in GBC.
\end{minipage}
\\[-1mm]\hline\\[-4mm]
(e)
&
\begin{minipage}{1.025\columnwidth}
\begin{lstlisting}[language=python]
k_neighbors_classifier = KNeighborsClassifier.customize_schema(
    n_neighbors={"type": "integer", "minimum": 1, "maximum": 40})
planned = (PCA | PolynomialFeatures) >> k_neighbors_classifier
\end{lstlisting}
\end{minipage}
&
\begin{minipage}{0.45\columnwidth}\raggedright\scriptsize
using tiny dataset folds, some values of \python{n_neighbors} exeed
the number of rows
\end{minipage}
&
\begin{minipage}{0.45\columnwidth}\raggedright\scriptsize
limit \python{n_neighbors} to be at most the number of rows in a fold
\end{minipage}
\\[-1mm]\hline\\[-4mm]
(f)
&
\begin{minipage}{1.025\columnwidth}
\begin{lstlisting}[language=python]
planned = PCA >> LogisticRegression
\end{lstlisting}
\end{minipage}
&
\begin{minipage}{0.45\columnwidth}\raggedright\scriptsize
\python{whiten=True} can cause NaN
\end{minipage}
&
\begin{minipage}{0.45\columnwidth}\raggedright\scriptsize
set \python{whiten=False} in \python{PCA}
\end{minipage}
\\[-1mm]\hline\\[-4mm]
(g)
&
\begin{minipage}{1.025\columnwidth}
\begin{lstlisting}[language=python]
int_1_63 = {"minimum": 1, "maximum": 63, "type": "integer"}
pca = PCA.customize_schema(n_components=int_1_63)
select_k_best = SelectKBest.customize_schema(k=int_1_63)
planned = pca >> select_k_best >> LogisticRegression
\end{lstlisting}
\end{minipage}
&
\begin{minipage}{0.45\columnwidth}\raggedright\scriptsize
pipeline instances where \python{n_components < k} pipe too few
columns to \python{SelectKBest}
\end{minipage}
&
\begin{minipage}{0.45\columnwidth}\raggedright\scriptsize
configure choices with non-overlapping ranges for
\python{n_components} and \python{k}
\end{minipage}
\\[-1mm]\hline\\[-4mm]
(h)
&
\begin{minipage}{1.025\columnwidth}
\begin{lstlisting}[language=python]
project = Project(columns={"type": "number"})
planned = (project | StandardScaler) >> LogisticRegression
\end{lstlisting}
\end{minipage}
&
\begin{minipage}{0.45\columnwidth}\raggedright\scriptsize
non-numbers for \python{StandardScaler}
\end{minipage}
&
\begin{minipage}{0.45\columnwidth}\raggedright\scriptsize
remove \python{StandardScaler} from choice
\end{minipage}
\\[-1mm]\hline\\[-4mm]
(i)
&
\begin{minipage}{1.025\columnwidth}
\begin{lstlisting}[language=python]
planned = LinearRegression | LogisticRegression
\end{lstlisting}
\end{minipage}
&
\begin{minipage}{0.45\columnwidth}\raggedright\scriptsize
non-regression dataset
\end{minipage}
&
\begin{minipage}{0.45\columnwidth}\raggedright\scriptsize
remove \python{LinearRegression}
\end{minipage}
\\[-1mm]\hline\\[-4mm]
(j)
&
\begin{minipage}{1.025\columnwidth}
\begin{lstlisting}[language=python]
project = Project(columns={"type":"number"})
imputer = SimpleImputer(strategy="most_frequent")
planned = project >> (imputer | NoOp) >> LogisticRegression
\end{lstlisting}
\end{minipage}
&
\begin{minipage}{0.45\columnwidth}\raggedright\scriptsize
missing values for \python{LogisticRegression}
\end{minipage}
&
\begin{minipage}{0.45\columnwidth}\raggedright\scriptsize
remove \python{NoOp} from choice (\python{|})
to ensure imputation
\end{minipage}
\\[-1mm]\hline\\[-4mm]
(k)
&
\begin{minipage}{1.025\columnwidth}
\begin{lstlisting}[language=python]
project = Project(columns=credit5_cats)
one_hot = OneHotEncoder(handle_unknown="ignore")
ordinal = OrdinalEncoder(handle_unknown="ignore")
planned = (project >> (one_hot | ordinal)
           >> StandardScaler >> LogisticRegression)
\end{lstlisting}
\end{minipage}
&
\begin{minipage}{0.45\columnwidth}\raggedright\scriptsize
similar to (b), but this time, it only fails
when \python{one_hot} is chosen
\end{minipage}
&
\begin{minipage}{0.45\columnwidth}\raggedright\scriptsize
either remove \python{one_hot} from choice or
set \python{with_mean=False} in \python{StandardScaler}
\end{minipage}
\\[-1mm]\hline\\[-4mm]
(l)
&
\begin{minipage}{1.025\columnwidth}
\begin{lstlisting}[language=python]
project = Project(columns=credit5_cats)
one_hot = OneHotEncoder(handle_unknown="ignore")
ordinal = OrdinalEncoder(handle_unknown="ignore")
sscaler = StandardScaler(with_mean=False)
planned = (project >> (one_hot | ordinal)
           >> (sscaler | RobustScaler) >> LogisticRegression)
\end{lstlisting}
\end{minipage}
&
\begin{minipage}{0.45\columnwidth}\raggedright\scriptsize
fails due to sparse data in pipeline instances that pick
\python{one_hot} for the encoder and \python{RobustScaler} for the
scaler
\end{minipage}
&
\begin{minipage}{0.45\columnwidth}\raggedright\scriptsize
either remove \python{one_hot} from encoder choice or
remove \python{RobustScaler} from scaler choice
\end{minipage}
\\[-1mm]\hline\\[-4mm]
(m)
&
\begin{minipage}{1.025\columnwidth}
\begin{lstlisting}[language=python]
project = Project(columns={"type": "number"})
planned = (project | NoOp) >> SimpleImputer >> DummyClassifier
\end{lstlisting}
\end{minipage}
&
\begin{minipage}{0.45\columnwidth}\raggedright\scriptsize
imputation strategy depends on upstream operator choice
\end{minipage}
&
\begin{minipage}{0.45\columnwidth}\raggedright\scriptsize
either remove \python{NoOp} or set \python{strategy="most_frequent"}
\end{minipage}
\\[-1mm]\hline\\[-4mm]
(n)
&
\begin{minipage}{1.025\columnwidth}
\begin{lstlisting}[language=python]
planned = GerryFairClassifier(**fairness_info) | LogisticRegression
\end{lstlisting}
\end{minipage}
&
\begin{minipage}{0.45\columnwidth}\raggedright\scriptsize
\python{GerryFairClassifier}, 3 classes
\end{minipage}
&
\begin{minipage}{0.45\columnwidth}\raggedright\scriptsize
remove \python{GerryFairClassifier}
\end{minipage}
\\[-1mm]\hline\\[-4mm]
(o)
&
\begin{minipage}{1.025\columnwidth}
\begin{lstlisting}[language=python]
decision_tree_classifier = DecisionTreeClassifier.customize_schema(
    max_features={"type": "integer", "minimum": 1, "maximum": 10})
planned = RobustScaler >> decision_tree_classifier
\end{lstlisting}
\end{minipage}
&
\begin{minipage}{0.45\columnwidth}\raggedright\scriptsize
not enough columns for \python{DecisionTreeClassifier}
\end{minipage}
&
\begin{minipage}{0.45\columnwidth}\raggedright\scriptsize
limit \python{max_features} to be at most the number of columns
\end{minipage}
\\[-1mm]\hline\\[-4mm]
(p)
&
\begin{minipage}{1.025\columnwidth}
\begin{lstlisting}[language=python]
unconstrained_lr = unconstrained_op(LogisticRegression)
planned = unconstrained_lr(dual=False)
\end{lstlisting}
\end{minipage}
&
\begin{minipage}{0.45\columnwidth}\raggedright\scriptsize
conditional hyperparameters \python{solver}, \python{multi_class}
\end{minipage}
&
\begin{minipage}{0.45\columnwidth}\raggedright\scriptsize
set \python{solver}$\,\neq\,$\python{"liblinear"} or \python{multi_class}$\,\neq\,$\python{"multinomial"}
\end{minipage}
\\[-1mm]\hline\\[-4mm]
(q)
&
\begin{minipage}{1.025\columnwidth}
\begin{lstlisting}[language=python]
unconstrained_pca = unconstrained_op(PCA)
planned = unconstrained_pca >> DummyClassifier
\end{lstlisting}
\end{minipage}
&
\begin{minipage}{0.45\columnwidth}\raggedright\scriptsize
conditional hyperparameters \python{n_components}, \python{svd_solver}
\end{minipage}
&
\begin{minipage}{0.45\columnwidth}\raggedright\scriptsize
set the hyperparameters to non-conflicting combinations
\end{minipage}
\\[-1mm]\hline\\[-4mm]
(r)
&
\begin{minipage}{1.025\columnwidth}
\begin{lstlisting}[language=python]
project = Project(columns={"type": "number"})
unconstrained_fa = unconstrained_op(FeatureAgglomeration)
planned = project >> unconstrained_fa >> LogisticRegression
\end{lstlisting}
\end{minipage}
&
\begin{minipage}{0.45\columnwidth}\raggedright\scriptsize
conditional hyperparameters \python{affinity}, \python{linkage}
\end{minipage}
&
\begin{minipage}{0.45\columnwidth}\raggedright\scriptsize
set \python{affinity="euclidian"} or
\python{linkage}$\,\neq\,$\python{"ward"}
\end{minipage}
\\[-1mm]\hline\\[-4mm]
(s)
&
\begin{minipage}{1.025\columnwidth}
\begin{lstlisting}[language=python]
unconstrained_gbm = unconstrained_op(LGBMClassifier).customize_schema(
  boosting_type={"enum":["gbdt","dart","goss","rf"],"default":"gbdt"})
planned = unconstrained_gbm
\end{lstlisting}
\end{minipage}
&
\begin{minipage}{0.45\columnwidth}\raggedright\scriptsize
conditional hyperparameters \python{boosting_type},
\python{subsample_freq}, and \python{subsample}
\end{minipage}
&
\begin{minipage}{0.45\columnwidth}\raggedright\scriptsize
set the hyperparameters to non-conflicting combinations
\end{minipage}
\\[-1mm]\hline\\[-4mm]
(t)
&
\begin{minipage}{1.025\columnwidth}
\begin{lstlisting}[language=python]
pca = unconstrained_op(PCA)(whiten=False)
one_hot = OneHotEncoder(handle_unknown="ignore")
pre_num = Project(columns={"type": "number"}) >> pca
pre_cat = Project(columns={"type": "string"}) >> one_hot
planned = (pre_num & pre_cat) >> ConcatFeatures >> LogisticRegression
\end{lstlisting}
\end{minipage}
&
\begin{minipage}{0.45\columnwidth}\raggedright\scriptsize
similar to (q), but this time, the problematic
\python{unconstrained_pca} is in a nested sub-pipeline
\end{minipage}
&
\begin{minipage}{0.45\columnwidth}\raggedright\scriptsize
set the hyperparameters of \python{unconstrained_pca} to
non-conflicting combinations
\end{minipage}
\end{tabular}
\caption{\label{fig:examples}Example planned pipelines with causes of failure and remediations.}
\end{table*}

\end{document}